**Contrasting magnetic behavior of fine particles of some Kondo Lattices**

K. Mukherjee, Kartik K. Iyer and E. V. Sampathkumaran*
*Tata Institute of Fundamental Research, Homi Bhabha Road, Colaba, Mumbai-400005 India*

**ABSTRACT**

We have investigated the magnetic behavior of ball-milled fine particles of well-known Kondo lattices, $CeAu_2Si_2$, $CePd_2Si_2$ and $CeAl_2$, by magnetization and heat-capacity studies in order to understand the magnetic behavior when the particle size is reduced. These compounds have been known to order antiferromagnetically in the bulk form near ($T_N=$) 10, 10 and 3.8 K respectively. We find that the features due to magnetic ordering get suppressed to temperatures below 1.8 K in the case of fine particles of ternary alloys, though trivalence of Ce as inferred from the effective moment remains unchanged. In contrast to this, in $CeAl_2$, there appears to be a marginal enhancement of $T_N$, when the particle size is reduced to less than a micron. These results can be consistently understood by proposing that there is relatively more 4f-localization as the particle size is reduced, resulting in weakening of exchange interaction strength.

## 1. Introduction

In the field of strongly correlated electron systems (SCES) containing 4f and 5f metals, the question of how the properties get modified when the particle size is reduced to micron or nano scale has not been sufficiently addressed in the literature. In this respect, we have initiated some work in recent years [1, 2] by synthesizing fine particles by high-energy ball-milling [3] in an atmosphere of toluene. It was noticed that $CeRu_2Si_2$, one of the most celebrated non-magnetic Kondo lattices in the bulk form, crystallizing in $ThCr_2Si_2$-type tetragonal structure, reveals features due to magnetic ordering around ($T_N$=) 8 K when the particle size is reduced, whereas isostructural $CeRh_2Si_2$, known to exhibit antiferromagnetic order near 36 K, loses magnetism. To consistently understand this, we argued that there could be a dramatic decrease in the 4f-conduction band coupling strength ($J$) for the ball-milled specimens with respect to that for bulk form, resulting in narrowing of 4f-band. We therefore consider it important to carry out similar investigations on other Ce compounds to place this inference on firm grounds. With this motivation, we have studied the magnetic behavior of ball-milled fine particles of three well-known magnetically ordering Ce compounds, two of which ($CeAu_2Si_2$ and $CePd_2Si_2$) are isostructural to above Ce compounds, whereas the third one ($CeAl_2$) forms in $MgCu_2$-type cubic structure. The milling conditions of these samples were kept the same to enable a reliable comparison of variations of the magnetic properties. At this juncture, it is worth mentioning that there were a few reports in the past on the magnetism of nano particles of rare-earth-based materials prepared by ball milling as well as those synthesized through a different route [4, 5].

The compounds $CeAu_2Si_2$, $CePd_2Si_2$ and $CeAl_2$, have been studied extensively in the bulk form [See, for example, References 6-20]. $CeAu_2Si_2$ and $CePd_2Si_2$ order antiferromagnetically at $T_N \sim 10K$ in the bulk form. From the high pressure studies [18], it was inferred that $J$ for the former is much weaker than in $CePd_2Si_2$. For instance, a pressure of 2.5 GPa is required to suppress magnetic ordering in the Pd compound, while for $CeAu_2Si_2$ the ordering persists under pressure up to 17 GPa [14]. The bulk form of $CeAl_2$ undergoes a complex antiferromagnetic ordering below 3.8 K [20]. We report here that the magnetic ordering tends to get suppressed in the ball-milled fine particles of $CeAu_2Si_2$ and $CePd_2Si_2$, while in $CeAl_2$, there appears to be an enhancement of magnetic ordering temperature, which could be consistent with the proposal that 4f electrons tend to get more localized in fine particles.

## 2. Experimental details

The polycrystalline ingots of $CeAu_2Si_2$, $CePd_2Si_2$ and $CeAl_2$ were prepared by arc melting stoichiometric amounts of high-purity (>99.9% in case of Ce and >99.99% for other elements) constituent elements in an arc furnace in an atmosphere of argon. The bulk samples thus obtained were characterized by x-ray diffraction (XRD) (Cu $K\alpha$); in addition, magnetic measurements on these ingots were carried out to ascertain the formation of proper phase before milling. The materials were then milled for 5 hours in a planetary ball-mill (Fritsch pulverisette-7 premium line) operating at a speed of 500 rpm in a medium of toluene. Zirconia vials and balls of 5mm diameter were used and the ball-to-material ratio was kept at 5:1. These vials and balls are made up of 96.4% of $ZrO_2$, and 3.2% of MgO, which ensures absence of contamination from magnetic impurities due to these milling parts. XRD patterns (shown in figure 1) confirm single-phase nature of all specimens (both for bulk and fine particles), the scanning electron microscopic (SEM, Nova NanoSEM600, FEI Company) studies established homogeneity, and the energy dispersive x-ray analysis (EDXA) confirmed atomic ratios of the constituent elements. In addition, a transmission electron microscope (TEM, Tecnai 200 kV) was also employed to characterize the milled specimens. The *dc* magnetization (*M*) measurements in the temperature (*T*) range 1.8-300 K for all specimens were carried out with the help of a commercial superconducting quantum interference device (Quantum Design). We have employed a commercial vibrating sample magnetometer (Oxford Instruments) to obtain isothermal magnetization behavior up to 120 kOe. In all magnetization measurements, the samples were cooled from a temperature (100 K) which is well-above the respective magnetic ordering temperatures, unless stated otherwise. We have also performed heat capacity (*C*) measurements on the fine particles employing a physical property measurements system as described in Ref. 2.

3. Results
3.1 X-ray diffraction patterns and particle sizes

The XRD patterns are shown in Fig. 1 for the milled ones along with the patterns for the bulk forms. No extra peaks are observed for both bulk and milled ones, which confirm the formation of proper phases. The lattice constants obtained by least-squares fit of d-spacings are also mentioned in this figure. From this figure, it is obvious that the XRD lines are broadened in the milled specimens. An idea of the average particle sizes was also inferred from the width of these lines (after subtracting instrumental line broadening) employing Debye-Scherrer formula. The values thus obtained are typically around 10, 18 and 36nm for $CeAu_2Si_2$, $CePd_2Si_2$ and $CeAl_2$ respectively. To correct for effects due to strain, well-known Williamson-Hall plots were not helpful as they were not found to be linear, possibly due to irregular shapes of milled specimens. In order to get a better idea about the particle size, we have employed SEM and TEM. According to the SEM images (not shown), the particles are agglomerated and therefore the sizes of the particles are actually much smaller. Some of the particles were isolated by ultrasonification in alcohol and the bright-field TEM images were obtained on these. These images shown in Fig. 2 (a, d, and g) reveal nano-particle formation for all compositions. The particle sizes obtained from TEM are of the same order of magnitude as that obtained from XRD for $CeAu_2Si_2$ and $CePd_2Si_2$. However for $CeAl_2$, the particle size obtained (with TEM in comparison to that from XRD) are more (~100 nm). We have also obtained selected area electron-diffraction patterns (see Fig. 2(b, e and h)) and all the diffraction rings are indexable for all the compounds under investigation, thereby confirming that the nanospecimens correspond to respective parent compounds and these are polycrystalline; the appearance of some bright spots along the diffraction rings reveals that the particles are highly textured. Furthermore high-resolution TEM images (HRTEM, see figures 2 c, f, i) show well defined lattice planes, thereby confirming that the nanospecimens are crystalline and not amorphous. Henceforth, the bulk will be called *B* while the milled specimens (fine particles) will be called *N* for each compound.

3.2 $CeAu_2Si_2$

Figure 3a shows the temperature dependence of magnetization for the zero-field-cooled (ZFC) and field-cooled (FC) conditions of the bulk and fine particles of $CeAu_2Si_2$. The magnetic ordering for the bulk form sets in at ~10K and there is no bifurcation between the ZFC and FC curves at this temperature, in agreement with the literature [15]. There is a feature near 6 K with a bifurcation of ZFC-FC curves at this temperature, which is suppressed at higher fields, say, at 5 kOe (see Fig.3b). This 6K-feature was reported in the single crystal as well [15]. For the fine particle form, it is seen that the above-mentioned feature due to antiferromagnetic ordering vanishes and the *M/H* versus *T* curve continuously rises with the decrease in temperature as though antiferromagnetic ordering is suppressed. No bifurcation was observed between the ZFC and FC curves for the fine particles as well at the onset of magnetic ordering. Inverse susceptibility ($\chi$) measured in a field of 5 kOe also varies essentially linearly below 50 K, as though the fine particles are essentially paramagnetic down to 1.8 K. The effective moment obtained above 100 K is ~ 2.4$\mu_B$/formula-unit, almost the same as that observed for the bulk form, establishing that Ce remains trivalent in the fine particles as well. The value of the paramagnetic Curie temperature ($\theta_p$) changes from ~ 23K for the bulk to ~ -30K for the fine particles. However, we do not wish to draw any conclusion for this compound on the basis of the changes in $\theta_p$, as the sign and the values have been known to be strongly anisotropic [6].

In order to render support to the inference on the destruction of antiferromagnetism in fine particles, isothermal magnetization as a function of magnetic-field shown in figures 3(c) and (d) for different temperatures are useful. As well-known, for the bulk form [15], a metamagnetic transition is observed at 50 kOe at 1.8 K, the signature of which is prevalent at 5K as well; for the fine particles, this metamagnetism is absent and there is a curvature with *M* varying smoothly with *H* as though there is a tendency to saturate at high fields typical of that expected for Ce-based paramagnets at this temperature, thereby implying that antiferromagnetism vanishes for the fine particles. The magnetic moment values at 120 kOe for the bulk and fine particles are nearly the same (about 0.8 $\mu_B$/Ce and 0.9 $\mu_B$/Ce at 1.8 K).

In order to understand the low temperature behavior better, we show the results of heat-capacity measurements in figure 3(e). For the bulk, a pronounced anomaly is seen at ~10 K, consistent with the onset of a magnetic phase transition. However, for the fine particles, though this feature is suppressed, the anomaly is not completely washed out and there is a feature with a feeble upturn appearing at about 8 K; the upturn is prominently seen in the *C/T* plot (figure 3f). Therefore, the possibility that a small fraction of Ce ions remaining magnetically ordered, however with a reduced ordering temperature (~8K), cannot be ruled out. The *C/T* versus $T^2$ plot is essentially linear in the range 12-20 K. The value of the linear term, $\gamma$ (obtained from $C = \gamma T + \beta T^3$), is negligible for the bulk and ~ 120 mJ/mol K$^2$ for the fine particles. It is not clear whether this implies strengthening of heavy-fermion character with decreasing particle size, as the estimation of electronic term from the heat-capacity data in this temperature range could be misleading due to possible interference from the Schottky peak due to crystal-field effects. Due to difficulties in reproducing fine particles of identical dimensions for the corresponding non-magnetic counterpart, we are not confident of obtaining 4f-contribution to heat-capacity and hence of any further inferences based on such derived data.

### 3.3 CePd$_2$Si$_2$

From figure 4a, it is clear that the peak due to magnetic ordering that appears around 10K in *M(T)* for the bulk form of this compound is suppressed in its fine particles, with the value of *M* undergoing monotonic increase with decreasing temperature, similar to the case of $CeAu_2Si_2$. There is no bifurcation of the ZFC-FC curves, measured in a field of 100 Oe as for the bulk form. The values of the effective moment obtained from the data measured in a field of 5 kOe in the Curie-Weiss regime are ~ 2.6 and ~ 2.5 $\mu_B$/formula-unit for bulk and fine particles respectively, typical of trivalency of Ce. Corresponding values of $\theta_p$ are about -65K and -49K for the bulk and fine particles respectively [19]. In the lower temperature range (10 to 40 K) as well, $\theta_p$ follows the same trend varying from ~ -44 to ~ -17K from *B* to *N*. It is seen that, even though magnetization varies linearly with field for the bulk $CePd_2Si_2$, in fine particles, the nature of the curve changes (figure 4b) characterizing it a paramagnet, similar to that of $CeAu_2Si_2$. Consistent with this, no hysteresis is observed in *M(H)* curve at 1.8K. [Incidentally, ferromagnetism was induced [21] with the reduction in particle size for $GdMn_2Ge_2$ and $TbMn_2Ge_2$]. Just as in *M(T)* data, a peak is observed in the plot of *C(T)* for the bulk (figure 4c), which vanishes for the fine particles, supporting the idea that the magnetic ordering is suppressed. For the fine particles, the value of $\gamma$ obtained from the range 5 to 10 K appears to be about 360mJ/mol $K^2$, establishing heavy-fermion behavior. A further upturn, though weak, is observed below 3 K in the *C/T* versus *T* plot (figure 4d). It is well-known that the Wilson ratio (*W*), defined as a dimensionless ratio of the zero-temperature magnetic susceptibility and the coefficient of the linear temperature term in the specific heat, gives an idea about magnetic correlations and its enhancement over unity is usually taken as a signature of increasing importance of magnetic interactions. The value of *W* was determined for the fine particles and it was found to be about 6, which is higher than that usually seen for heavy fermion systems (close to unity). This suggests that either magnetic short range correlation persists or there are traces of Ce ions remaining magnetic, resulting in somewhat higher values of magnetic susceptibility at 1.8 K. Incidentally, the value of *W* is found to be the same for the fine particles of $CeAu_2Si_2$ as well.

**3.4 $CeAl_2$**

For this compound, as seen in figure 5a, the peak known for the bulk form around 4 K in *M/H* plot persists for the milled specimens. A careful comparison of the two curves in figure 5a reveals that the peak is replaced by a broader peak, but shifted to a higher temperature (around 5 K) for the fine particles, in contrast to the two cases discussed above; in addition there is an upturn at lower temperatures which is absent for the bulk specimen. The broadening of the feature could also be attributable to defects introduced by ball-milling. Also for the milled sample, under FC condition, no peak is observed as compared to its *B* counterpart and *M/H* keeps increasing with decreasing temperature. The overall changes in features could imply a change in the nature of magnetism due to milling.

In addition, we note an interesting finding for the fine particles: That is, there is a bifurcation of the ZFC-FC curves, obtained in a field of 100 Oe, below 20 K. This raises a question whether there is any other magnetic anomaly at a temperature much higher than $T_N$.

The effective moment for the fine particles in the Curie-Weiss region (above 100 K) is typical of trivalent Ce. The value of high-temperature $\theta_p$ is essentially unaltered (~ -26 K), as though the single-ion Kondo temperature behavior for fully degenerate state is also not unaffected much. In the low temperature region of 20-50K, $\theta_p$ gets modified due to crystal-field effects, with the value changing from ~ -18K (for bulk) to ~ -9K (for fine particles). While the *M* vs. *H* curves (figure 5b) show no hysteresis for both the forms, the spin-reorientation-like feature that appears for the bulk form around 20-40 kOe [20] vanishes for the fine particles; there appears to be two linear regions intersecting near 30 kOe. Thus, the nature of the curvature for the bulk and for the fine particles are different, establishing a change in the low-temperature magnetic character.

The appearance of a λ-anomaly in *C(T)* (shown in figure 5c) also supports the proposal of persistence of magnetic ordering when the particle-size is reduced, though the anomaly is relatively more broadened presumably due to defects introduced by milling. However, as in the *M(T)* data (figure 5a), the peak around 3.5 K for the bulk is shifted to 5 K for the fine particles, confirming that the magnetic ordering temperature is enhanced in the fine particles for this compound. An interesting observation we have made is that the milling appears to introduce a ferromagnetic component, as an application of a magnetic field marginally causes an upward shift of the peak in *C(T)* (inset of figure 5c). Since a strong peak in the heat-capacity data is representative of the bulk of a material, we conclude that the ferromagnetism does not arise from Ce ions at the grain boundaries. The observation of a ferromagnetic component also explains the inference from the magnetization data with respect to the modification of magnetic structure for the fine particles.

**4.    Discussions**

A common observation made for the milled specimens is that the magnitude of $\theta_p$ at low temperatures decreases with a reduction in particle size, which is clearly evident at least for $CePd_2Si_2$ and $CeAl_2$. The sign however remains negative and the Ce valence as inferred from the effective moment also does not change (that is, trivalent). These findings viewed together imply that for the crystal-field-split ground state, the Kondo interaction strength is decreased for these particles, which means more localization of 4f orbital. This 4f-localisation apparently pushes the two ternary compounds to the left of the peak in Doniach's magnetic phase diagram [22] to the extent that the magnetic ordering temperature is suppressed. In the case of $CeAl_2$, a small application of external pressure lowers the Néel temperature for the bulk form [16], and this compound has been known to lie at the right side of the peak in Doniach's magnetic phase diagram. Therefore, an increase of magnetic ordering temperature in the milled specimen is expected for milled specimens due to the proposed increase of 4f-localisation due to milling. This is found to be the case experimentally. These ideas are consistent with the conjecture in Ref. 2.

Finally, Han et al [4] reported that the nano particles of $CeAl_2$ actually lose antiferromagnetism. Therefore, the physical properties described in the present article imply that the particles studied in this article are relatively bigger in size as evidenced by the TEM image for this sample. The fact that the magnetic transition temperature can be dependent on particle size was actually demonstrated for an isostructural compound $CePt_2$ [5].

**5.  Conclusion**

The magnetic properties of fine particles of well known Ce-based Kondo lattices have been addressed. The long range magnetic ordering seen for the bulk form of $CeAu_2Si_2$ and $CePd_2Si_2$ is suppressed in the fine particle form. However, for $CeAl_2$, there appears to be a marginal enhancement of magnetic ordering temperature. This observation could be consistently understood by the proposal that there is a tendency for 4f-localization with decreasing particle size. Finally, we would like to mention that the nano particles of CePd$_2$Si$_2$ prepared [23] by laser evaporation do not reveal any magnetic transition down to 1.8 K, similar to the findings on the ball-milled particles reported in this article.

**Acknowledgement**
We thank Y. Y. Chen for sharing the results on nano particles of CePd$_2$Si$_2$ synthesized by laser ablation method before publication, R. D. Bapat for TEM measurements and S. D. Das for his participation during the initial phase of this work.

*Corresponding author: sampath@tifr.res.in

**References**
[1] Sitikantha D. Das, S Narayana Jammalamadaka, Kartik K. Iyer, and E. V. Sampathkumaran Phys. Rev. B 80 (2009) 024401; K. Mukherjee, Kartik K. Iyer and E. V. Sampathkumaran J. Phys.: Condens. Matter 22 (2010) 295603.
[2] E. V. Sampathkumaran, K. Mukherjee, Kartik K. Iyer, Niharika Mohapatra and Sitikantha D. Das   J. Phys.: Condens Matter 23 (2011) 094209.
[3] C Suryanarayana Prog. Mater. Sci.46 (2001) 1**.**
[4] G. F. Zhou, H. Bakker Phys. Rev. B 52 (1995) 9437; I. W. Modder, H. Bakker Phys. Rev. B 58 (1998) 14479; M. A. Morales, D. S. Williams, P. M. Shand, C. Stark, T. M. Pekarek, L. P. Yue, V. Petkov, D. L. Leslie-Pelecky Phys. Rev.B 70 (2004) 184407; D. P. Rojas, L. F. Barquin, J. R. Fernandez, J. I. Espeso, J. C. G. Sal J. Phys.: Condens. Matter 19 (2007) 186214; D. P. Rojas, L. F. Barquin, J. I. Espeso, J. R. Fernandez, J. Chaboy Phys. Rev. B 78 (2008) 094412.
[5] S. W. Han, C. H. Booth, E. D.  Bauer, P. H. Huang, Y. Y. Chen and J. M. Lawrence. Phys. Rev. Lett. 97 (2006) 097204; C. L. Dong, Y. Y. Chen, C. L. Chen, J. H. Guo and C. L. Chang  J. Magn. Magn. Mater 304 (2006) e22; Y. Y. Chen, P. H. Huang, M. N. Ou, C. R. Wang, Y. D. Yao, T. K. Lee, M. Y. Ho, J. M. Lawrence and C. H. Booth  Phys. Rev. Lett. 98 (2007) 157206; J. S. Kim, G. R. Stewart, K. Samver Phys. Rev. B 79 (2009) 165119; Y. Y. Chen, Y. D. Yao, T. K. Lee, C. Tse, W. C. Liu, H. C. Chang, K. Y. Lin, Y. S. Lin, Z. C. Wang and W. H. Li.  Chinese J. Phys. 36 (1998) 2.
[6] Y. Ota,  Kiyohiro Sugiyama,  Yuichiro Miyauchi, Yuji Takeda,  Yasunori Nakano, Yusuke Doi,  Keisuke Katayama, Nguyen D Dung, Tatsuma D  Matsuda, Yoshinori Haga, Koichi Kindo,  Tetsuya Takeuchi,  Masayuki Hagiwara,  Rikio Settai and Yoshichika Ōnuki.  J. Phy. Soc. Jpn 78 (2009) 034714.
[7] C. S. Garde and J. Ray  J Phys.: Condens. Matter. 6 (1994) 8585.
[8] V. Vildosola, A. M. Llois, M. Alouani Phys. Rev. B 71 (2005) 184420.
[9] R. D. Parks, B. Reihl, N. Maartensson and F. Steglich  Phys. Rev. B 27 (1983) 6052.
[10] N. Kernavanois, S. Raymond, E. Ressouche,  B. Grenier, J. Flouquet and P. Lejay Phys. Rev. B 71 (2005) 064404.
[11] I. Sheikin, A. Gröger, S. Raymond, D. Jaccard, D. Aoki, H. Harima and J. Flouquet J. Magn. Magn. Mater 272 (2004) e33.
[12] H. Miyagawa,  Gendo Oomi,  Masashi Ohashi, Isamu Satoh, Takemi Komatsubara, Masato Hedo,  Yoshiya Uwatoko.  Phys. Rev. B 78 (2008) 064403.

[13] F. Steglich  Physica B 378 (2006) 7.
[14] P. Link and D. Jaccard  Physica B 230 (1997) 31.
[15] A. S. Sefat, Andriy M. Palasyuk, Sergey L.Bud'ko, John D. Corbett and Paul C.Canfield  J Solid State Chem 181 (2008) 282.
[16] M. C. Croft, R. P. Guertin, L. C. Kupferberg and R. D. Parks Phys. Rev. B 20 (1979) 2073.
[17] R. Movshovich, T. Graf, D. Mandrus, M. F. Hundley, J. D. Thompson, R. A. Fisher, N. E. Phillips and J. L. Smith  Physica B 223-224 (1996) 126.
[18] J. D. Thompson, R. D. Parks and H. J. Borges  J. Magn. Magn. Mater 54-57 (1986) 377.
[19] E. V. Sampathkumaran, Y. Nakazawa, M. Ishikawa and R. Vijayaraghavan Phys. Rev. B 40 (1989) 11452.
[20] B. Barbara, M. F. Rossignol, J. X. Boucherle, J. Schweizer and J. L. Buevoz.  J. Appl. Phys. 50 (1979) 2300.
[21] K. Mukherjee, Kartik K. Iyer and E. V. Sampathkumaran Euro. Phys. Letts. 90 (2010) 17007.
[22] S. Doniach Physica B 1977; 91: 231.
[23] Y. Y.Chen et al (private communication).

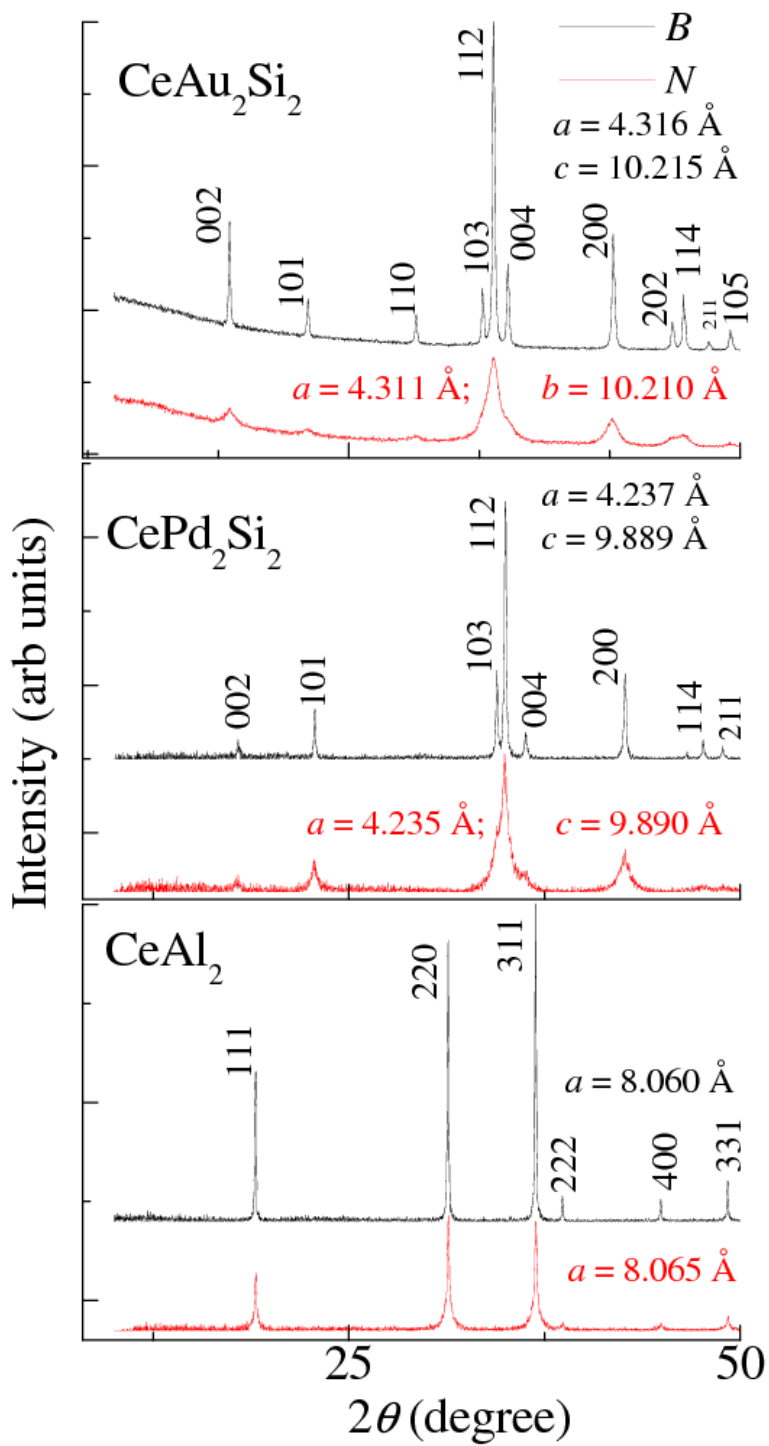

**Fig. 1.** X-ray diffraction patterns (Cu K$\alpha$) for the bulk and the milled specimens. In all cases, the Miller indices are given. The lattice constants (±0.004 Å), '$a$' and '$c$' are also included. The broadening of lines in the milled specimen is visible in the respective curves.

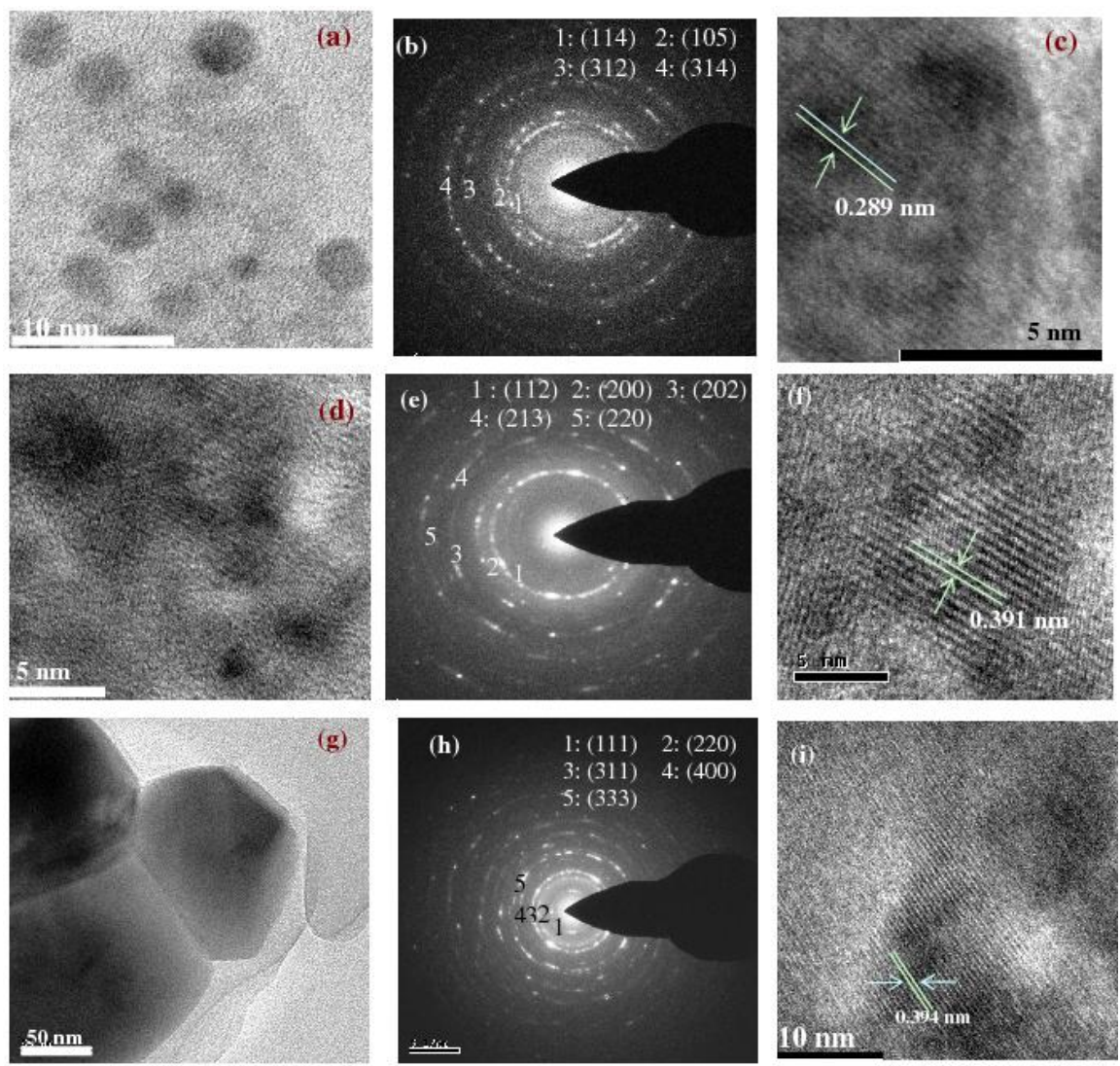

**Fig. 2.** Transmission Electron Microscope (TEM) images, selected area diffraction pattern obtained by TEM with indexing of the innermost diffraction rings and HRTEM pictures (with spacings in units of Å marked), for the milled specimens of (a,b,c) $CeAu_2Si_2$, (d,e,f) $CePd_2Si_2$ and (g,h,i) $CeAl_2$.

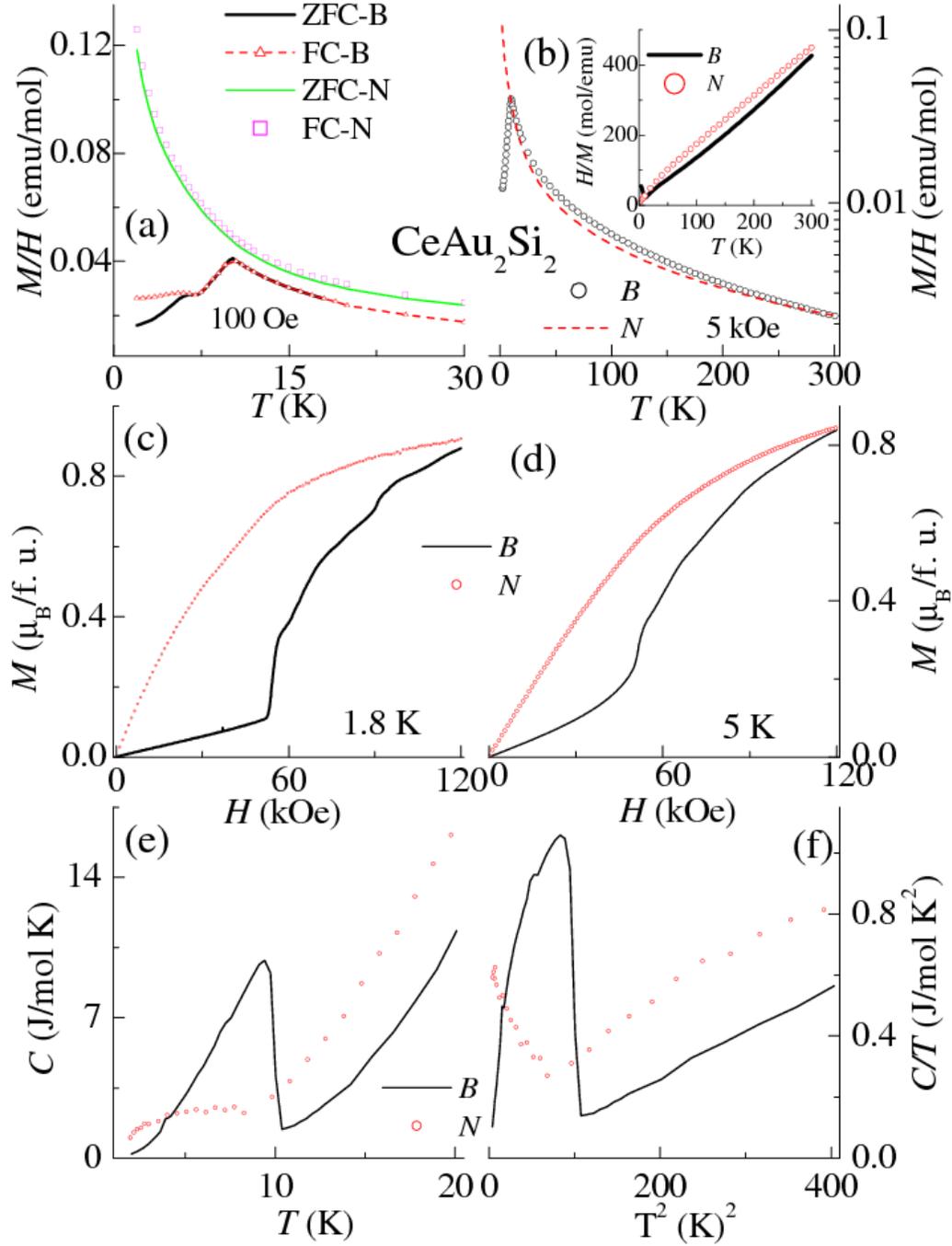

**Fig. 3.** For CeAu$_2$Si$_2$, **(a)** temperature ($T$) dependence (below 60 K) of magnetization ($M$) divided by magnetic field ($H$) for zero-field-cooled and field-cooled specimens of both bulk ($B$) and fine particles ($N$), and **(b)** $T$-dependence of $M/H$ at 5 kOe. Inset: $H/M$ values measured in 5 kOe for $B$ and $N$ are plotted as a function of $T$. **(c)** and **(d)** Isothermal magnetization at 1.8 and 5K respectively, **(e)** heat capacity ($C$) and **(f)** $C/T$ as a function of $T$ for both $B$ and $N$ in zero field. For the sake of clarity, for some curves, the lines (though data points) only are shown.

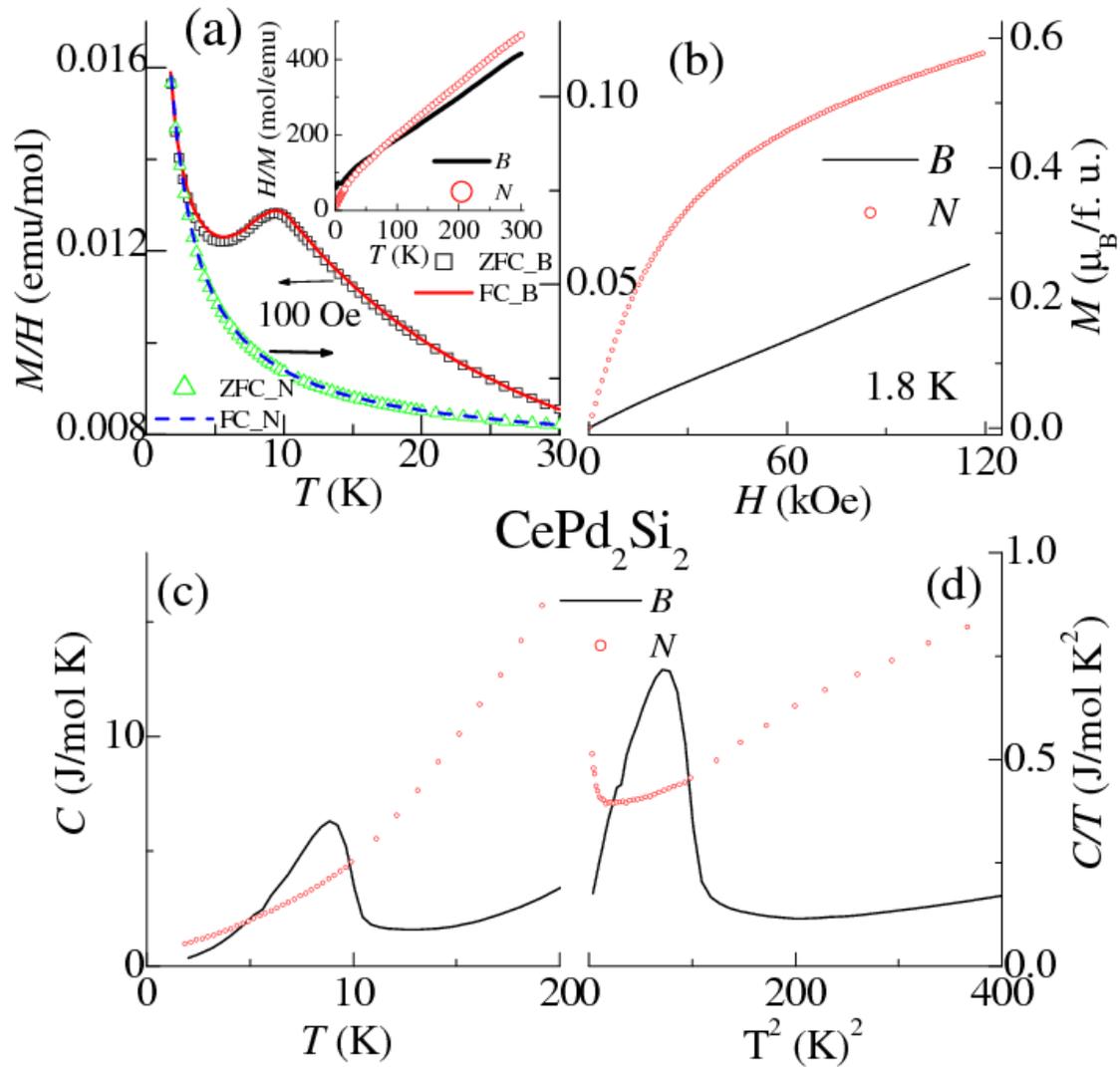

**Fig. 4.** For bulk and milled samples of $CePd_2Si_2$, **(a)** temperature ($T$) dependence (below 30 K) of magnetization ($M$) divided by magnetic field ($H$) for zero-field-cooled and field-cooled conditions; inset: $H/M$ values measured in 5 kOe are plotted as a function of $T$. **(b)** Isothermal magnetization at 1.8K, **(c)** heat capacity ($C$), and **(d)** $C/T$ as a function of temperature in zero field. For the sake of clarity, for some curves, the lines (though data points) only are shown.

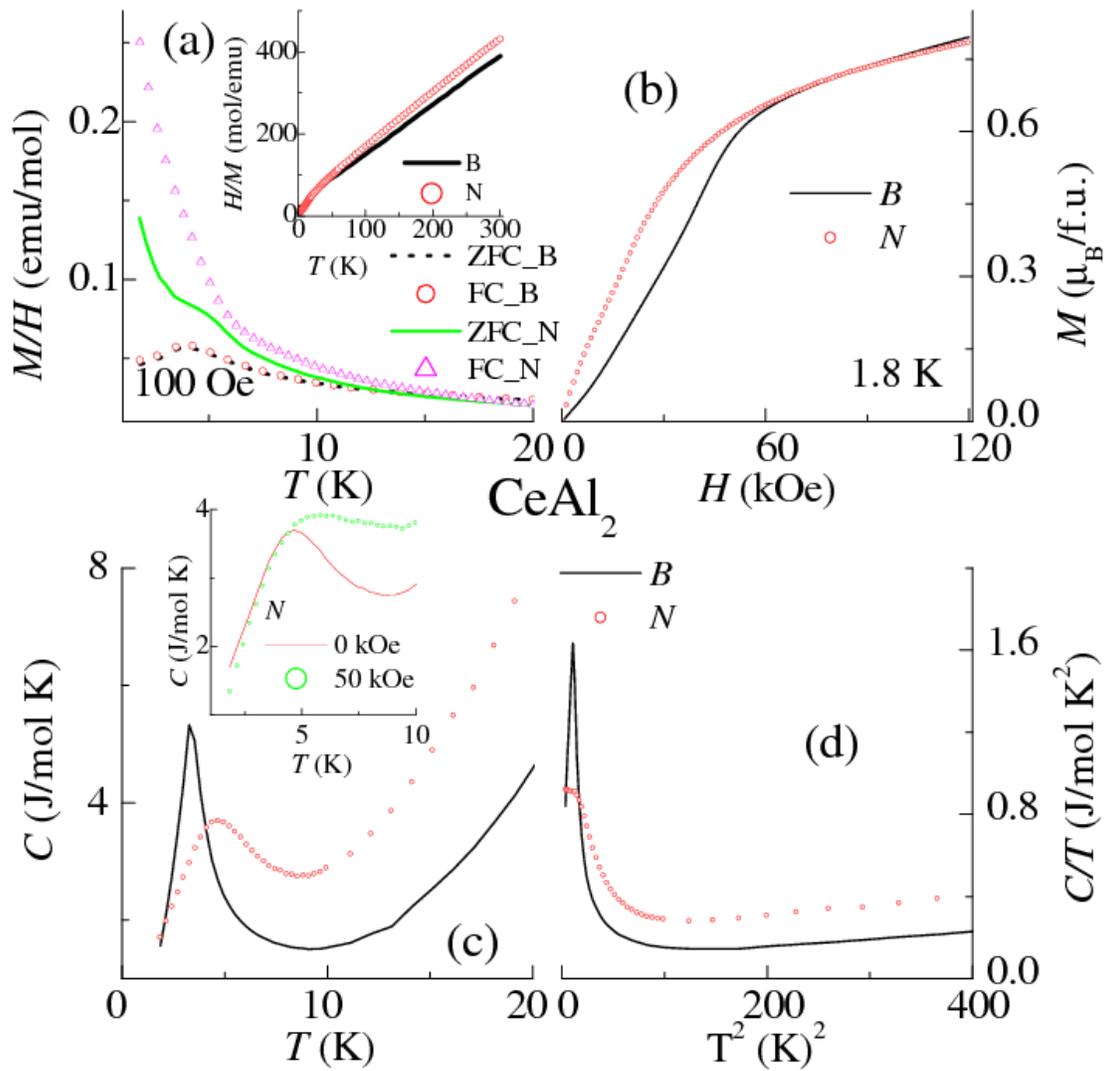

**Fig. 5.** For bulk and milled samples of CeAl$_2$, **(a)** temperature ($T$) dependence (below 30 K) of magnetization ($M$) divided by magnetic field ($H$) for zero field cooled and field cooled specimens; Inset: $H/M$ values measured in 5 kOe are plotted as a function of $T$. **(b)** Isothermal magnetization at 1.8K, **(c)** heat capacity ($C$), and **(d)** $C/T$ as a function of temperature in zero field. In inset of **(c)**, $C$ as a function of T for milled sample in 0 and 50 kOe fields is plotted. For the sake of clarity, for some curves, the lines (though data points) only are shown